\documentclass[aps,prl,floatfix,showpacs,twocolumn]{revtex4}
\usepackage{amssymb,amsmath}
\usepackage{stmaryrd}
\usepackage{graphicx}
\usepackage{epsfig}
\usepackage{psfrag}
\newcommand{\bm}{\boldsymbol}

\begin{document}

\hsize\textwidth\columnwidth\hsize\csname@twocolumnfalse\endcsname
\title{Magneto-optical and Magneto-electric Effects of Topological Insulators in Quantizing Magnetic Fields} 

\author{Wang-Kong Tse}
\author{A.~H. MacDonald}
\affiliation{Department of Physics, University of Texas, Austin, Texas 78712, USA}

\begin{abstract}
We develop a theory of the magneto-optical and magneto-electric
properties of a topological insulator thin film in the presence of a 
quantizing external magnetic field.
We find that low-frequency magneto-optical properties
depend only on 
the sum of top and bottom surface Dirac-cone 
filling factors $\nu_{\mathrm{T}}$ and $\nu_{\mathrm{B}}$,  
whereas the low-frequency magneto-electric response depends only on the
difference. 
The Faraday rotation is quantized in integer multiples of the fine
structure constant and the Kerr effect exhibits a $\pi/2$
rotation. Strongly enhanced cyclotron-resonance features appear at 
higher frequencies that are sensitive to the filling factors of both surfaces.
When the product of the bulk conductivity and the film thickness in
$e^2/h$ units is small compared to $\alpha$, 
magneto-optical properties are only weakly dependent on accidental doping in the 
interior of the film.
\end{abstract}
\maketitle

\noindent
{\em Introduction}---
Topological insulators (TIs) are a recently identified \cite{KaneHasan} new class of materials.  Three-dimensional TIs have insulating bulks
and metallic surfaces with an odd number of Dirac cones that are responsible for most unique TI properties.
Angle-resolved photoemission spectroscopy (ARPES) experiments have established that several 
strongly spin-orbit coupled materials \cite{ARPES}
exhibit TI properties.  
In this paper we develop a theory of the magneto-optical and
magneto-electric properties of TI thin films in the presence of a perpendicular 
external magnetic field. 

Our work is motivated in part by 
potential advantages of magneto-optical over transport \cite{Transport} characterization in
isolating TI surface properties from 
bulk contamination due to unintended doping.  
Since Landau level (LL) quantization of the TI's surface Dirac cones has recently been
established by STM experiments \cite{STM}, it should be possible to detect 
surface quantum Hall effects optically, even when parallel bulk conduction is present.
In the quantum Hall regime, we find that  
the low-frequency Faraday effect is \textit{quantized} in integer multiples of the fine structure constant, 
while the Kerr effect displays the same \cite{Tse_TI} giant $\pi/2$ rotation relative to the 
incident polarization direction that appears when time reversal is broken by exchange coupling.
At higher frequencies, we find strong cyclotron resonance
features in both Faraday and Kerr spectra.    

One goal of our work is to clarify 
how the magneto-electric 
effects peculiar to TIs \cite{SCZhangRev,Vanderbilt,Tse_TI,VanderbiltII,QiBulkSubstrate} 
are reflected in their thin film magneto-optical properties.  
We show that low-frequency TI  
magneto-optical response in the quantum Hall regime depends on the 
sum of top and bottom surface filling factors, whereas the magneto-electric response
of film polarization to an external magnetic field depends on the filling factor 
difference.  We argue that coupling between electric and magnetic fields in the presence
of a TI material is most usefully regarded as a property of its surfaces, not of its bulk. 

\noindent
\textit{Low-frequency Magneto-electric and Magneto-optical Response}---
The response of a TI thin film to an external magnetic field is
dominated by \cite{addargument} 
its Dirac-cone surface states.  When the Dirac cone quantum Hall effect 
is well developed, its Hall conductivity $\sigma_{xy}$ has quantized plateau values with half-odd-integer values in $e^2/h$ units.  
The longitudinal resistivity vanishes on the Hall plateaus, but is non-zero on the {\em risers} between
Hall plateaus. The risers between the $(n-1/2)(e^2/h)$ and
$(n+1/2)(e^2/h)$ 
plateaus occur near integer values of the filling factor $\nu =n \in
\mathbb{Z}$.
The Dirac cone's quantum Hall effect has so far been cleanly observed
\cite{grapheneHall} only in two-dimensional graphene, 
in which four separate Dirac cones conduct in parallel. 

The Streda \cite{streda} formula, 
\begin{equation}
\label{streda}
\sigma_{xy} = ec (\partial N/\partial B) = e c  (\partial M/\partial\mu),
\end{equation} 
which is valid at Hall plateau centers, implies a relationship between 
surface conductivities and the magneto-electric response of a TI thin film.
In Eq.~(\ref{streda}) $N$ is the 
two-dimensional electron density,
$B$ the external magnetic field, $M$ the orbital magnetization, and $\mu$ the 
chemical potential. 
We define the electric polarization $P$ per unit volume of a TI thin film 
in terms of the difference between the surface charge densities
accumulated on the top ($\mathrm{T}$) and 
bottom $(\mathrm{B})$ surfaces $P = e (N_{\mathrm{T}}-N_{\mathrm{B}})/{2}$.
It follows that for plateaus characterized by $\nu_{\mathrm{T}}$ and $\nu_{\mathrm{B}}$
the magneto-electric susceptibility 
\begin{equation} 
\chi_{\mathrm{ME}}= 4\pi \partial P/\partial B = 
(\nu_{\mathrm{T}}-\nu_{\mathrm{B}})\alpha
\label{alphaME},
\end{equation} 
depends only on the filling factor difference between surfaces 
($\alpha=e^2/\hbar c$ is the fine structure constant). 
Recent experiments have demonstrated \cite{TIdopingexp} 
the possibility of tuning the surface carrier densities systematically by surface doping or gating. 
At a fixed magnetic field, the top and bottom surfaces need
not be on the same Hall plateau.
Provided that the bulk resistance is sufficiently large, $\chi_{\mathrm{ME}}$ can 
be measured by contacting top and bottom surfaces separately and 
detecting voltages induced by magnetic field variation.
If sample quality can be improved sufficiently, TI thin films might be useful as 
magnetic field sensors. 

The TI's surface quantum Hall effect can also be probed by Faraday and Kerr angle measurements. 
In the low frequency regime the electromagnetic wavelength is
much longer than the film thickness. 
Faraday's and Amp\`ere's Laws then imply that the electric field is 
spatially constant across the film, while the magnetic field 
jumps by a value proportional to the current 
integrated across the TI film.  
Allowing for a bulk conductivity $\Sigma$ due 
to unintended doping \cite{Transport} and assuming a small bulk Hall angle
we find that the low-frequency quantum-Hall-regime Faraday and Kerr angles
depend on the TI surface properties only through $\nu_{\mathrm{T}}+\nu_{\mathrm{B}}$:
\begin{eqnarray}
\tan\theta_{\mathrm{F}} &=&
\frac{\left(\nu_{\mathrm{T}}+\nu_{\mathrm{B}}\right)\alpha}{1+2\pi\Sigma
d/c}, \label{Faraw0}
\\
\tan\theta_{\mathrm{K}} &=&
\frac{4\left(\nu_{\mathrm{T}}+\nu_{\mathrm{B}}\right)\alpha}{1-(1+4\pi\Sigma
  d/c)^2-[2\left(\nu_{\mathrm{T}}+\nu_{\mathrm{B}}\right)\alpha]^2}. \label{Kerrw0}
\end{eqnarray}
\begin{figure}
  \includegraphics[width=8.7cm,angle=0]{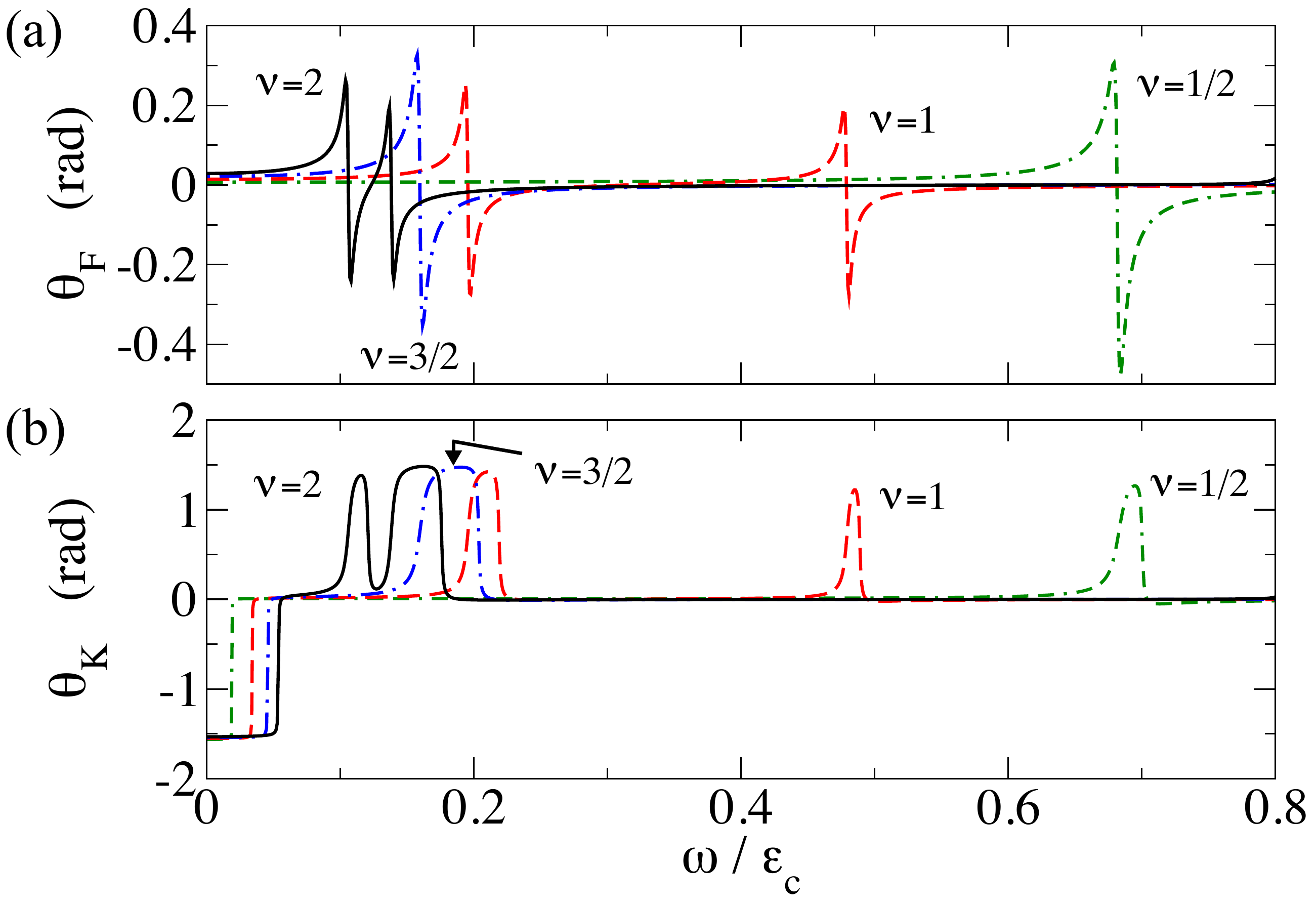}
\caption{(Color online). (a). Faraday rotation $\theta_{\mathrm{F}}$ versus frequency $\omega/\varepsilon_{\mathrm{c}}$ at 
equal filling factors on both surfaces $\nu = 1/2$ (green), $\nu = 1$ (red), $\nu = 3/2$ (blue), and $\nu = 2$ (black). 
The densities on both surfaces are $N_{\mathrm{T,B}}
=5\times10^{11}\,\mathrm{cm}^{-2}$. (For this density the filling factor 
is given by $\nu=20.81/B\,[{\rm Tesla}]$.) We choose a 
$30\,\mathrm{nm}$-thick Bi$_2$Se$_3$ film as a prime example of TIs 
with a large bulk band gap $E_{\mathrm{g}} = 0.35\,\mathrm{eV}$. The 
Fermi velocity is $v = 5\times 10^5\,\mathrm{cm}^{-2}$ and the 
dielectric constant $\epsilon = 29$.  (b). Kerr rotation $\theta_{\mathrm{K}}$
versus frequency at the same filling
factors.} 
\label{fig_thetaF}
\end{figure}
The bulk carriers enter
as an \textit{effective} longitudinal surface conductivity 
$\Sigma d$, where $d$ is the film thickness. 
When the bulk conductivity is sufficiently small so that ${\Sigma d}/{(e^2/h)}
\lesssim \alpha$, 
the Kerr angle 
exhibits a universal full-quarter rotation $\theta_{\mathrm{K}} \simeq
\pm\pi/2$. 
When ${\Sigma d}/{(e^2/h)} \ll 1/\alpha$, the Faraday angle is quantized in integer multiples of the fine
structure constant
\begin{equation}
\theta_{\mathrm{F}} \simeq (\nu_{\mathrm{T}}+\nu_{\mathrm{B}})\alpha. \label{Fara_TF}
\end{equation}
Given the rapid progress in TI film quality, these regimes 
should be within reach experimentally; 
in particular, the less stringent condition 
for Faraday angle quantization 
should be currently accessible. 
Since the magneto-electric polarizability Eq.~(\ref{alphaME}) yields
the filling factor difference, and the Faraday angle
Eq.~(\ref{Fara_TF}) yields the filling factor sum, measurement of both
quantities could allow the filling factors $\nu_{\mathrm{T,B}}$
to be extracted individually. 

\noindent
\textit{Dirac-Cone ac Conductivity}---
Outside of the long-wavelength limit, TI thin-film magneto-optical properties
depend on the finite-frequency Dirac-cone conductivity which we now evaluate microscopically.
The high-frequency signal consists of resonances at inter-LL
transition frequences. 
We neglect optical phonon contributions to the conductivity which are not expected to be 
significantly dependent on magnetic field strength.
In an external magnetic field the Dirac-cone 
Hamiltonians for the top (T) and bottom (B) surfaces are 
$H = (-1)^L \; [ v\bm{\tau}\cdot(-i\nabla+e\bm{A}/c) + V/2] + \Delta\tau_z$,
where $\bm{\tau}$ is the spin Pauli matrix vector, $\bm{A} = (0,Bx)$ is the vector potential, 
$\Delta = g\mu_{\mathrm{B}}B/2$ is the Zeeman coupling, $V$ accounts for a possible 
potential difference between top and bottom surfaces due to doping or external gates, and $L = 0,1$ 
for the top ($0$) and bottom ($1$) surfaces.  The LLs are labeled by integers $n$ and 
for $n \ne 0$ have eigenenergies (relative to the Dirac point energies $(-1)^LV/2$)
$\varepsilon_n = \mathrm{sgn}(n)\sqrt{{2v^2}\vert n
  \vert /{\ell_B^2}+\Delta^2}$, 
where $\ell_B = \sqrt{c/e\vert B\vert}$ is the magnetic length.
In the $n=0$ LL spins are aligned with the perpendicular field 
and $\varepsilon_0 = -\Delta$.
For convenience we rewrite the LL index as $n = sm$, where $m = 0, 1,
2, \cdots N_{\mathrm{c}}$ and 
$s = \pm 1$ for electron-like and hole-like LLs.
$N_{\mathrm{c}} \simeq
\ell_B^2(\varepsilon_{\mathrm{c}}^2-\Delta^2)/2v^2$ 
is the largest LL index
with an energy smaller than the ultraviolet cut-off $\varepsilon_{\mathrm{c}}$.
We choose $\varepsilon_{\mathrm{c}} = E_{\mathrm{g}}/2$ where $E_{\mathrm{g}}$
is the bulk band gap.

Using the Kubo formalism we find that in the quantum Hall regime ($\Omega_B\tau
\gg 1$, where $1/\tau$ the quasiparticle lifetime broadening and
$\Omega_B = v/\ell_B$ is a characteristic frequency typical of the LLs
spacing)  
the conductivity in 
$e^2/\hbar = \alpha c $ units
is given by
\begin{eqnarray}
&&\sigma_{\alpha\beta}(\omega) = \label{Cond} \\
&&\frac{v^2}{2\pi \ell_B^2}\mathrm{sgn}(B)\sum_{m = 0}^{N_{\mathrm{c}}-1}\sum_{s,s' = \pm 1}
\frac{f_{sm}-f_{s'(m+1)}}{\varepsilon_{sm}-\varepsilon_{s'(m+1)}} \; \Gamma_{\alpha\beta}^{s,s'}(m,\omega), \nonumber
\end{eqnarray}
Here $\alpha,\beta = \{x,y\}$, 
$f_{sm}$ is a Fermi factor,
and
\begin{eqnarray}
&&\Gamma_{\left\{\begin{subarray}{l} xx \\ xy \end{subarray}\right\}}^{s,s'}(m,\omega) = 
-\left\{\begin{array}{c} i \\ 1 \end{array}
\right\}\mathcal{C}_{\uparrow s'(m+1)}^2\mathcal{C}_{\downarrow
  sm}^2 
\label{function_f} \\
&&\left(\frac{1}{\omega-\varepsilon_{sm}+\varepsilon_{s'(m+1)}+i/2\tau}\pm\frac{1}{\omega+\varepsilon_{sm}-\varepsilon_{s'(m+1)}+i/2\tau}\right). \nonumber
\end{eqnarray}
In Eq.(~\ref{function_f}), the LL eigenspinors are $\mathcal{C}_{\uparrow 0} 
= 0$, $\mathcal{C}_{\downarrow 0} = 1/\sqrt{2}$, and for $m \ne 0$ 
$\mathcal{C}_{\uparrow sm} = s\sqrt{\varepsilon_{m}+s \Delta}/\sqrt{2\varepsilon_{m}}$, and 
 $\mathcal{C}_{\downarrow sm} = \sqrt{\varepsilon_{m}-s
   \Delta}/\sqrt{2\varepsilon_{m}}$. 
Eqs.~(\ref{Cond})-(\ref{function_f}) express 
$\sigma_{\alpha\beta}$ as a sum over interband and intraband dipole-allowed 
transitions which satisfy $\vert n'\vert-\vert n\vert =\pm 1$.
In the $\omega=0$ limit Eq.~(\ref{Cond}) yields 
correct half-quantized plateau values for the Hall conductivity.

\noindent
\textit{TI Thin Film Cyclotron Resonance---}
Resonances at transition energies are resonsible for dips and peaks 
in transmission and reflection spectra respectively, and 
give rise to associated characteristic features in Faraday and Kerr spectra. 
Fig.~\ref{fig_thetaF}a shows the frequency dependence of the Faraday rotation for 
several different filling factors.  These results are obtained 
by combining the {\em ac} 
Dirac-cone conductivity with the
scattering matrix formalism described in Ref.~\cite{Tse_TI}. 
We find that $\theta_{\mathrm{F}}$ exhibits sharp 
cyclotron resonance peaks and changes sign near each allowed
transition frequency. At half-odd-integer filling factors there is a 
single dipole-allowed intraband transition; at other filling factors there are two allowed 
resonances at different frequencies associated with transitions into and out of the
partially filled LL.
This behavior is completely different from that of an ordinary
two-dimensional electron system 
where all dipole-allowed transitions have the same energy.
Interband transitions from hole-like to electron-like LLs 
are weaker and appear in the same energy range as the intraband
transitions  
only if the field is weak and the filling factor is small.
\begin{figure}
  \includegraphics[width=9cm,angle=0]{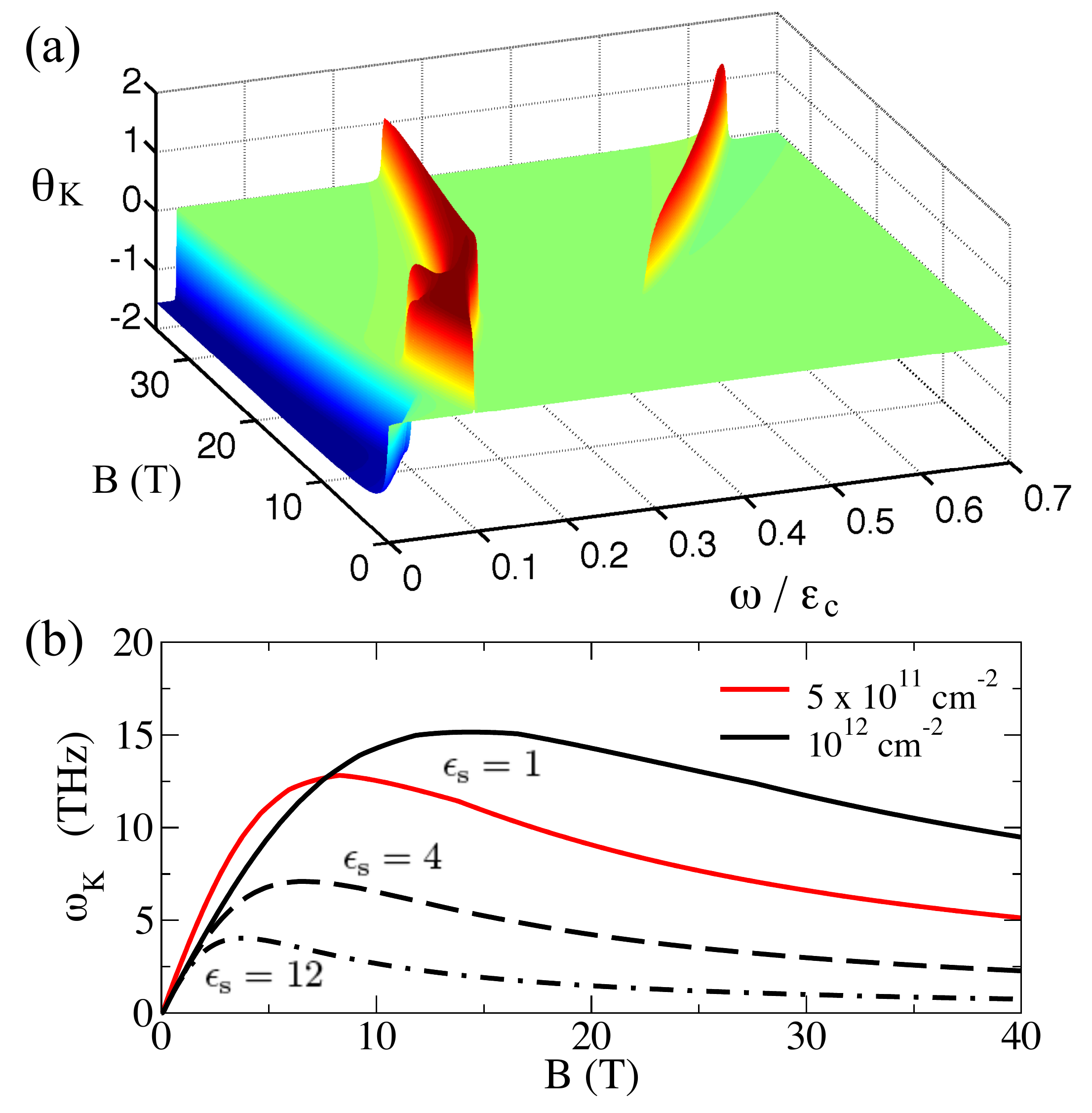}
\caption{(Color online) (a). Kerr angle $\theta_{\mathrm{K}}$ as a function of frequency 
  $\omega/\varepsilon_{\mathrm{c}}$ and magnetic field $B$ for surface densities 
$N_{\mathrm{T,B}} = 5\times10^{11}\,\mathrm{cm}^{-2}$.
Cyclotron resonance features corresponding to particular transitions 
are allowed when the initial Landau level is at least partially occupied and the final
Landau level is at least partially empty. In the weak-field semiclassical transport regime,
$\sigma_{xy} \propto B$ and $\theta_{\mathrm{K}}$ vanishes as $B \to
0$.
(b) Kerr frequency $\omega_{\mathrm{K}}$ (see text) as a function of 
  magnetic field for $N_{\mathrm{T,B}} = 10^{12}\,\mathrm{cm}^{-2}$ (dark/black) 
for free-standing $30\,\mathrm{nm}$-thick Bi$_2$Se$_3$
  ($\epsilon_{\mathrm{s}} = 1$),  
with SiO$_2$ substrate 
  ($\epsilon_{\mathrm{s}} = 4$), and  
Si substrate ($\epsilon_{\mathrm{s}} = 12$). Grey/red line shows the
  free-standing case with $5\times10^{11}\,\mathrm{cm}^{-2}$. The substrate thickness
  $d_{\mathrm{s}} = 1\,\mu\mathrm{m}$.}
\label{fig_thetaK}
\end{figure}

The Kerr angle $\theta_{\mathrm{K}}$ also shows dramatic 
enhancement at the cyclotron resonance frequencies
(Fig.~\ref{fig_thetaF}b).
Unlike the low-frequency giant 
Kerr effect, 
the resonant peaks which correspond to absorptive LL transitions 
are 
non-universal. 
We note that Zeeman coupling shifts the $0^{\mathrm{th}}$ LL away from the Dirac 
point and therefore breaks the electron-hole (e-h) symmetry of the LL
spectrum. 
This implies that the $n = -1\to0$ and $0\to1$ transition frequencies, 
$\omega_{0\to1} = \sqrt{2v^2/l_B^2+\Delta^2}+\Delta$ and $\omega_{-1\to 0} = \sqrt{2v^2/l_B^2+\Delta^2}-\Delta$, 
become different. 
In general, $\theta_{\mathrm{K}}$ is an odd function 
of filling factor $\nu$ if e-h symmetry is intact. 
When the surface filling factors $\nu_{\mathrm{T}} = -\nu_{\mathrm{B}}
= 1/2$ are opposite, e-h asymmetry implies that cyclotron resonance peaks of
opposite sign appear in $\theta_{\mathrm{K}}$ at the two transition
frequencies. The Kerr effect thus provides a smoking gun to detect
the absence of e-h symmetry.

In Fig.~\ref{fig_thetaK}a we plot the Kerr angle $\theta_{\mathrm{K}}$ as a function
of frequency and magnetic field for fixed surface carrier
densities.  As in the exchange coupling case \cite{Tse_TI}, the giant Kerr effect survives up to 
a relatively large frequency which we refer to as the Kerr frequency: 
\begin{equation}
\omega_{\mathrm{K}} = 
\frac{2\pi\sigma_{xy}^{\mathcal{R}}(0)/c}{\left[\left(\epsilon-\mu\right)d+\left(\epsilon_{\mathrm{s}}-\mu_{\mathrm{s}}\right)d_{\mathrm{s}}\right]-2\pi{\sigma_{xx}^{\mathcal{I}}}'(0)/c}. \label{freqwin}
\end{equation}
Here $\sigma_{xx},\sigma_{xy}$ are the total surface conductivities, 
$\epsilon, \mu, d$ are the dielectric constant, permeability ($\simeq 1$), and thickness of
the TI film, $\epsilon_{\mathrm{s}}, \mu_{\mathrm{s}},
d_{\mathrm{s}}$ are the corresponding substrate values, and $'$ in
$\sigma_{xx}^{\mathcal{I}}$ denotes a frequency derivative. 
It follows from Eq.~(\ref{freqwin}) that
$\omega_{\mathrm{K}}$ is inversely proportional to the optical
thickness of the TI film and the substrate. 
Fig.~\ref{fig_thetaK}b illustrates the effect of varying $B$ on 
$\omega_{\mathrm{K}}$. There exists an optimal value of 
$B$ for which $\omega_{\mathrm{K}}$ is in the terahertz range.  
High $\omega_{\mathrm{K}}$ values are most readily achieved on `low-$\kappa$' substrate materials like SiO$_2$ or using 
free-standing films suitable for optical studies \cite{Geim}.
$\omega_{\mathrm{K}}$ can also be increased by surface doping. 

\noindent
\textit{Discussion---} 
TI's have interesting magneto-electric and magneto-optical properties
when time reversal symmetry (TRS) is broken to open up a gap in its Dirac-cone surface states.  
TRS can in principle be broken 
by exchange coupling to an insulating ferromagnet, 
although it is not yet established that a sufficiently strong coupling 
can be achieved in practice.  
The circumstance discussed here in which a perpendicular magnetic field is applied to a TI thin
film provides an experimentally simpler and phenomenologically richer 
method for producing TI thin films with weak TRS breaking.
The special case in which the filling factor $\nu = 1/2$ on both surfaces yields 
the same surface Hall conductivities as in the exchange-coupled case
and therefore the same \cite{Tse_TI} low-frequency Kerr and Faraday angle results.
In the magnetic field case, however, strong magneto-optical and magneto-electric effects occur over a broad range of 
separately controllable carrier densities on both surfaces.  
The top and bottom surface Dirac cone filling factors, $\nu_{\mathrm{T}}$ and $\nu_{\mathrm{B}}$, 
can be identified \cite{Potemski} from the ratios of magneto-optical resonance
frequencies and from the patterns they produce in Faraday or Kerr spectra. 
Bulk conduction has a quantitative influence on Faraday and Kerr
spectra only when $\Sigma d/(e^2/h)$, the bulk contribution to the dimensionless effective surface conductivity,  
is larger than about
$1/\alpha$ and $\alpha$ respectively.  In the quantum Hall regime,
the Kerr angle is large whenever the filling factors sum to a non-zero value and 
are away from one of the integer values at which longitudinal resistance peaks occur.
On quantum Hall plateaus magneto-optical properties depend only on the sum of 
individual surface filling factors, whereas the magneto-electric response of film 
polarization to field depends only on the difference.  

The influence of a bulk TI material on electromagnetic fields can be 
represented \cite{SCZhangRev} 
by introducing an $\bm{E} \cdot \bm{B}$ term with coefficient $\alpha_{\mathrm{ME}}$ in the electromagnetic Lagrangian. 
The microscopic Poisson and Amp\`ere
equations 
then take \cite{Wilczek,dissipationless} the form  
\begin{eqnarray} 
&&\nabla\cdot\bm{E} = 4\pi\rho-\nabla\alpha_{\mathrm{ME}}\cdot\bm{B},
\nonumber \\
&&\nabla\times\bm{B}-(1/c)\partial\bm{E}/\partial t = ({4\pi}/{c}) \bm{J}+ \nabla\alpha_{\mathrm{ME}}\times\bm{E}.
\label{eulerlagrange} 
\end{eqnarray} 
When both surfaces are on Hall plateaus, Eqs.~(\ref{eulerlagrange}) with $\rho$ and $\bm{J}$ set to zero 
is a valid course-grained description of the interface 
provided that we set 
\begin{equation} 
\label{dalphadz} 
{\mathrm{d} \alpha_{\mathrm{ME}}}/{\mathrm{d}z} = ({4\pi}/{c})\sigma_{xy}\,\delta(z) = 2 \alpha \nu\,\delta(z).
\end{equation}  
Because \cite{KaneHasan,VanderbiltII} 
$\nu$ can change by an integer 
without altering the bulk, 
only $\alpha_{\mathrm{ME}}$ modulo $2 \alpha$ characterizes the bulk material. 
(When the bulk $\alpha_{\mathrm{ME}}$ is mapped to the interval $[0,2\alpha]$,
only $\alpha_{\mathrm{ME}}=\alpha$ is consistent with 
time-reversal invariance.)  
In the thin film geometry normally employed in experiment, however,
a bulk value of $\alpha_{\mathrm{ME}}$ modulo $2 \alpha$ is not sufficient to predict the result of a magneto-optical or 
magneto-electric measurement \cite{Geometry} because it does not specify 
the Hall plateau index. 
Moreover, in real samples with non-zero disorder
the Dirac-cone surface Hall conductivities will \cite{Sinitsyn} be strongly suppressed
when the external magnetic field is weak. 
The values of $\alpha_{\mathrm{ME}}$ required to correctly 
reproduce the low-frequency magneto-optical signal will therefore become small and vanish 
with the external magnetic field.
In the stronger-field quantum Hall regime, the appropriate magneto-electric constant for vacuum,
TI thin-film, and substrate regions can be obtained by integrating Eq.~(\ref{dalphadz}) to 
obtain $\alpha_{\mathrm{ME}}=0$ in the upper vacuum, $\alpha_{\mathrm{ME}}= 2 \alpha
\nu_{\mathrm{T}}$ in the TI thin film, and $\alpha_{\mathrm{ME}}= 2 \alpha (\nu_{\mathrm{T}} +
\nu_{\mathrm{B}})$ in the substrate and lower vacuum. In both cases the 
appropriate values of $\alpha_{\mathrm{ME}}$ depend critically on
surface properties.  Although $\alpha_{\mathrm{ME}}\,\mathrm{mod}(2\alpha)$ 
is indeed a bulk property of a disorder-free TI, its value does not normally provide 
enough information to predict either magneto-optical properties or
magneto-electric response.

This work was supported by the Welch grant  F1473 and by DOE
grant DE-FG03-02ER45985. The authors acknowledge helpful discussions
with Kenneth Burch, Dennis Drew, Zhong Fang, David Vanderbilt and 
Shou-Cheng Zhang.


\begin{thebibliography}{20}
\bibitem{addargument}  The influence of non-magnetic matter on electromagnetic waves is 
completely captured by its frequency dependent conductivity tensor.
At frequencies well below the TI gap an external magnetic field has a negligible influence on the 
bulk conductivity.  When carriers are present in the bulk due to unintended doping, magnetic field dependence is 
still expected to be weak when disorder is strong.
\bibitem{Geometry} In addition, the coupling $\alpha_{\mathrm{ME}}$ is
  equivalent 
to 
the magneto-electric susceptibility $\chi_{ME}$ only for TI
geometries with one simply connected surface when $\chi_{ME} = 2\nu_{\mathrm{surface}}\alpha$ such as a sphere or
semi-infinite slab, but not for thin film.
\end{thebibliography}
\end{document}